\pdfoutput=1

\documentclass[review]{elsarticle}
\usepackage[margin=3cm]{geometry}

\usepackage{graphicx}
\usepackage{amssymb}
\usepackage{tabularx}
\usepackage{booktabs} 
\usepackage{siunitx}
\usepackage{placeins}
\usepackage{color}
\usepackage{soul}
\usepackage{textcomp}
\usepackage{hyperref}

\hypersetup{
	colorlinks   = true, %Colours links instead of ugly boxes
	urlcolor     = blue, %Colour for external hyperlinks
	linkcolor    = blue, %Colour of internal links
	citecolor   = red %Colour of citations
}

\biboptions{sort&compress}

\journal{Journal of the Mechanical Behavior of Biomedical Materials}

\begin{document}

	\begin{frontmatter}
		
		\title{A technique for improving dispersion within polymer-glass composites using polymer precipitation}

		\author[label1]{Reece N. Oosterbeek\corref{cor1}\fnref{fn1}}
		\ead{rno23@cantab.ac.uk}
		
		\author[label2]{Xiang C. Zhang}
		\author[label1]{Serena M. Best}
		
		\author[label1]{Ruth E. Cameron  \corref{cor1}}
		\ead{rec11@cam.ac.uk}
		
		\cortext[cor1]{Corresponding author}
		\fntext[fn1]{Present address: Department of Mechanical Engineering, Imperial College London, London, SW7 2AZ, United Kingdom; r.oosterbeek@imperial.ac.uk}
		
		\address[label1]{Cambridge Centre for Medical Materials, Department of Materials Science and Metallurgy, University of Cambridge, Cambridge, United Kingdom}
		\address[label2]{Lucideon Ltd, Queens Road, Penkhull, Stoke-on-Trent, United Kingdom}
		
		\begin{abstract}
		
			Particulate reinforcement of polymeric matrices is a powerful technique for tailoring the mechanical and degradation properties of bioresorbable implant materials. Dispersion of inorganic particles is critical to achieving optimal properties, however established techniques such as twin-screw extrusion or solvent casting can have significant drawbacks including excessive thermal degradation or particle agglomeration. We present a facile method for production of polymer-inorganic composites that reduces the time at elevated temperature and the time available for particle agglomeration. Glass slurry was added to a dissolved PLLA solution, and ethanol was added to precipitate polymer onto the glass particles. Characterisation of parts formed by subsequent micro-injection moulding of composite precipitate revealed a significant reduction in agglomeration, with $d_{0.9}$ reduced from 170 to 43 \textmu m. This drastically improved the ductility ($\varepsilon_{B}$) from 7\% to 120\%, without loss of strength or stiffness. The method is versatile and applicable to a wide range of polymer and filler materials.\\
			
		\end{abstract}
		
		
	\end{frontmatter}

	%%%%%%%%%%%%%%%%%%%%%%%%%%%%%%%%%%%%%%%%%%%%%%%%%%%%%%%%%%%%%%%%%%%%%%%%%%%%%%%%%%%%%%%%%%%%%%%%%%%%%%%

	\section{Introduction}
	
	Bioresorbable polymers such as poly-\textsc{L}-lactide (PLLA) are a crucial and ubiquitous tool for development of modern bioresorbable medical implant devices \cite{Perale2016, Oosterbeek2019}. However due to a range of functional limitations, the polymer matrix is often supplemented with incorporation of bioactive fillers. Inorganic phases such as hydroxyapatite, tricalcium phosphate, and bioactive glasses have all been investigated as composite components, and are a convenient way to provide mechanical reinforcement while also regulating degradation kinetics via a buffering effect \cite{Naik2017, Yang2009, Ahmed2011, Felfel2015}. The properties of these filler particles, and in particular their dispersion and agglomeration within the matrix, can have a significant effect on the properties of the composite, and can be highly dependent on the processing method and conditions. Dispersion of small particles within the polymer matrix to create the polymer composite can be challenging and requires careful method design. 
	
	Twin-screw or compounding extrusion (also known as melt blending) is a widely used technique for incorporating a particulate filler material into a polymer composite. This requires the use of a heated extruder/compounder, and uses the rotation of the screw(s) inside the extruder to mix the filler and polymer together \cite{Wang1994, Wang1998, Wilberforce2011a}. This has the advantage of being a scalable method that is already widely used industrially, but does require specialised equipment, and involves extended thermal cycling at high temperature combined with shear forces which can degrade the molecular weight of the polymer \cite{Odell1986}, leading to loss of mechanical properties and degradation resistance \cite{Boccaccini2016, Blaker2010}.
	
	Solvent-based methods offer a versatile and simple alternative method of producing polymer glass composites, however the use of hazardous solvents is a clear drawback. In addition, it can be difficult to remove all traces of residual solvent from the composite, which can be harmful to cells or tissue, as well as having a plasticising effect on the mechanical properties \cite{Kim2006}. Although the filler component can be well dispersed into the slurry, particle agglomeration can occur during solvent removal, depending on the drying speed and conditions, leading to inferior properties \cite{Boccaccini2016}.
	
	In this work we present a method that allows for straightforward fabrication of well-dispersed polymer-inorganic composites, by modification of established solvent-casting techniques. The microstructures of the composites are characterised, and the resulting mechanical properties are assessed.

	%%%%%%%%%%%%%%%%%%%%%%%%%%%%%%%%%%%%%%%%%%%%%%%%%%%%%%%%%%%%%%%%%%%%%%%%%%%%%%%%%%%%%%%%%%%%%%%%%%%%%%%

	\section{Materials and Methods}

	\subsection{Materials}
	Phosphate glass, with nominal composition (P\textsubscript{2}O\textsubscript{5})\textsubscript{45}(CaO)\textsubscript{45}(Na\textsubscript{2}O)\textsubscript{10} was produced as described elsewhere \cite{Oosterbeek2020}. Briefly, precursors (Na$_{2}$CO$_{3}$, CaCO$_{3}$, and NH$_{4}$H$_{2}$PO$_{4}$) were melted in a vitreous silica crucible in a kiln (SBSC-1500L, Kilns and Furnaces Ltd., Stoke-on-Trent, UK) at 1250\textcelsius\ and quenched by pouring onto a cooled steel plate. The glass frit was then crushed with a mortar and pestle, to fit through a 3.15 mm aperture sieve. PLLA (Ingeo 2500 HP) was supplied by Natureworks LLC, USA, DCM (dichloromethane) was supplied by Merck KGaA, Germany, and acetone and ethanol were purchased from Sigma Aldrich, UK.
	
	\subsection{Glass milling}
	Coarse milling was carried out using a Fritsch Pulverisette 6 planetary ball mill. Glass frit ($\sim$170 g) was added to the grinding bowl, along with 25 zirconia balls (20 mm diameter), before milling at 550 rpm for 5 min. A Netzsch PE 075 attritor mill was used for fine milling, with 400 g of milling media (0.5 mm ZrO\textsubscript{2} beads, Netzsch ZetaBeads Plus 0.5), 40 g glass powder, and 40 g acetone, milled for 60 min at 1500 rpm. After milling, the glass and ZrO\textsubscript{2} media were rinsed with acetone through a 200 \textmu m sieve, to remove the media and produce a dilute glass/acetone slurry. A small sample of the slurry was pipetted into a petri dish and left to dry to measure the slurry concentration ($\sim$ 60 mg mL\textsuperscript{-1}).
	
	\subsection{Composite fabrication}
\begin{figure}[h]
		\centering
		\includegraphics[width=11cm]{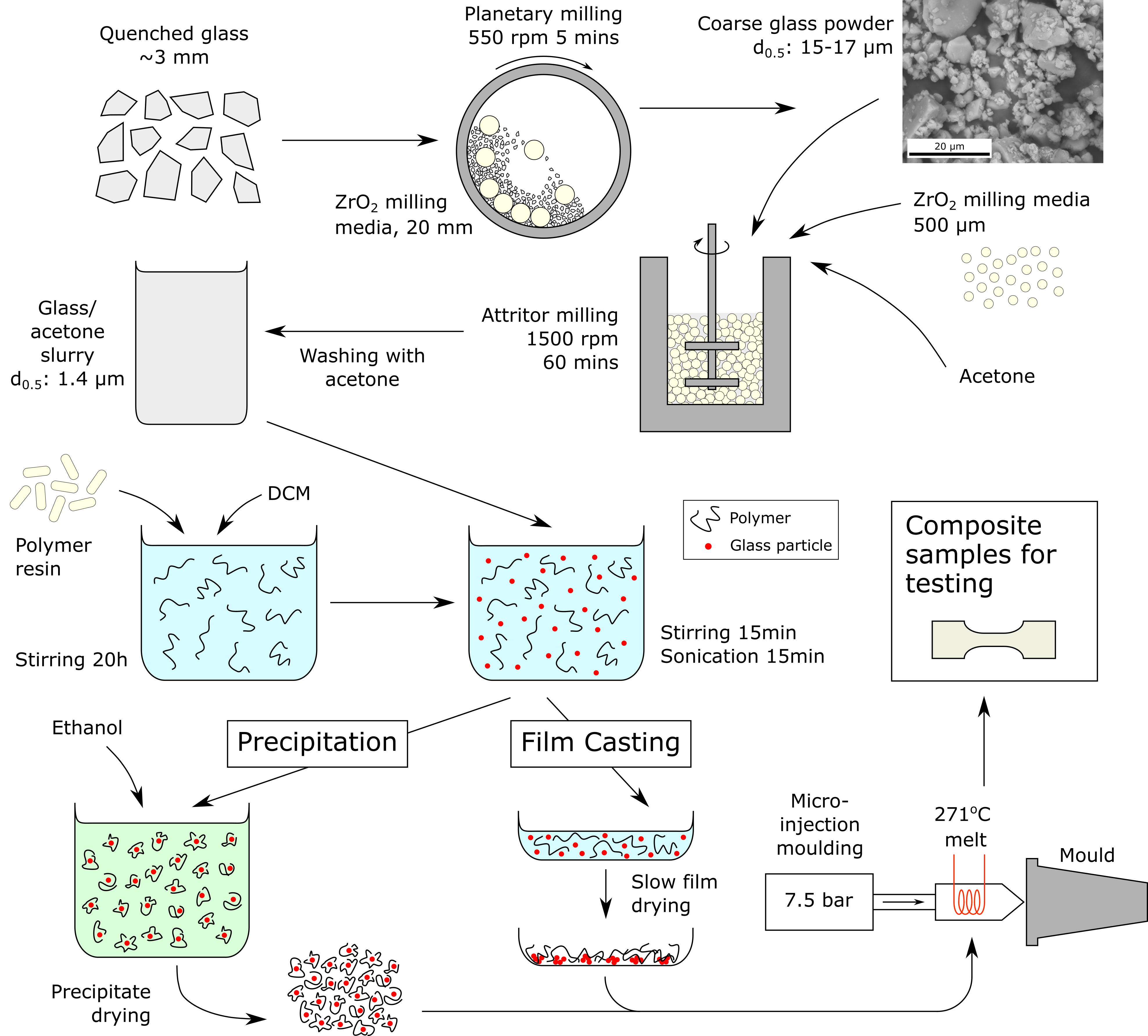}
		\caption{Schematic of composite production workflow, showing glass milling stages, composite production by precipitation or film casting, and then micro-injection moulding of composite parts.}
		\label{fig:CompWorkflow}
	\end{figure}

	Pure PLLA was dissolved in DCM (0.1 g mL\textsuperscript{-1}), and glass slurry was added to give the desired polymer/glass ratio, in this case 30 wt.\% glass. The dissolved polymer/glass slurry mixture was then stirred for 15 min and sonicated for 15 min at room temperature. To produce composite films this mixture was cast into petri dishes to dry in ambient conditions. After 24 hours drying in ambient conditions, films were dried under vacuum at 50\textdegree C for 10 days to remove any residual solvent. Polymer glass composites were also fabricated via a novel precipitation method. After stirring and sonication of the dissolved polymer/glass slurry mixture, ethanol was added while stirring, in a 3:1 ethanol/DCM ratio. This reduced the solvent power of the liquid, causing the polymer to precipitate onto the glass particles. This resulted in a clear solution with a solid composite precipitate, which was poured into a steel drying tray for the solvent to evaporate. The precipitate was then dried under vacuum at 50\textdegree C for 10 days to remove any residual solvent. Composite precipitates or cut films were then processed into dumbbell (19 $\times$ 5.5 $\times$ 0.6 mm) samples using micro-injection moulding (IM 5.5, Xplore Instruments BV, The Netherlands) and custom-made moulds in ambient conditions. 0.7-0.8 g of the polymer or composite material was loaded into the barrel, which was then heated to the minimum (nominal) melt temperature required for complete mould filling and uniform sample appearance (determined from previous trials to be 271\textdegree C). Once the desired temperature was reached, compressed air pressure of 7.5 bar was applied to the plunger to inject melted material into the room temperature mould. This pressure was held for 60 s to fill the mould and minimise shrinkage during cooling. Once cooled to room temperature, the sample was removed from the mould. The overall procedure is summarised in Figure \ref{fig:CompWorkflow}.
	
	\subsection{Characterisation}
	Glass particle size was measured using a Malvern Mastersizer 2000, with Hydro MU 2000A (Malvern Instruments Ltd, UK). The refractive index of the glasses was taken as 1.543, with absorption coefficient of 0.02 \cite{VenkateswaraRao2014}, and samples were dispersed in acetone. Pumping speed of 2000 rpm, and 60 s ultrasonication were used, and a 60 s delay was used after sonication to allow dissipation of any bubbles introduced, before beginning the analysis. Five measurements were carried out for each sample. Reported median diameters ($d_{0.5}$) are based on the particle volume distribution. SEM (Scanning Electron Microscopy) of glass particles was carried out using a CamScan MX2600 FEGSEM, with backscatter detector, using a 20 kV accelerating voltage. Prior to analysis coarse glass powder was sprinkled onto carbon tape, or fine glass slurry was dropped onto a glass slide and allowed to evaporate, before coating with 10 - 20 nm of Au/Pd, using an Emitech K550 sputter coater (20 mA deposition current for 4 min, under an argon atmosphere). Composite samples were imaged using an FEI Nova NanoSEM, using an accelerating voltage of 5 kV and secondary electron detector. Samples were prepared by cryo-fracturing and then sputter coating as above. X-ray micro-computed tomography (\textmu CT) analysis was carried out on composites using a Skyscan 1172 system (Bruker). X-ray voltage and current of 59 kV and 167 \textmu A were used, with 0.2\textdegree\ rotation steps, 2.58 s acquisition time, averaged over two frames. The pixel size was set at 1.49 \textmu m. Image projections were reconstructed into 3D datasets using the NRecon software (Bruker). 3D datasets were analysed using CTAn software (Bruker); global thresholding and despeckling were applied before individual object analysis was used to generate a size distribution of filler particles.
	
	\subsection{Mechanical testing}
	Tensile testing was carried out using a 1ST Benchtop Tester (Tinius Olsen Ltd, UK) with a 1 kN load cell, using miniature vice grips (HT54, Tinius Olsen Ltd, UK), under a constant elongation rate of 1 mm min\textsuperscript{-1}. Dumbbell samples (5 mm gauge length) were tested in simulated body conditions (immersed in deionised water at 37\textdegree C) using a Saline Test Tank with Heater (Tinius Olsen Ltd, UK). After loading samples into the grips and immersing them in water, they were left for approximately 10 min for the temperature to equilibrate. Strain was measured using a video extensometer and custom-built LabVIEW software. Yield strength ($\sigma_{y}$) was taken as the 0.2\% offset yield point, and the elastic modulus ($E$) was calculated from the linear region of the stress-strain curve before yield.

	%%%%%%%%%%%%%%%%%%%%%%%%%%%%%%%%%%%%%%%%%%%%%%%%%%%%%%%%%%%%%%%%%%%%%%%%%%%%%%%%%%%%%%%%%%%%%%%%%%%%%%%

	\section{Results and Discussion}

	\subsection{Glass particle characterisation}
\begin{figure}[h]
		\centering
		\includegraphics[width=11cm]{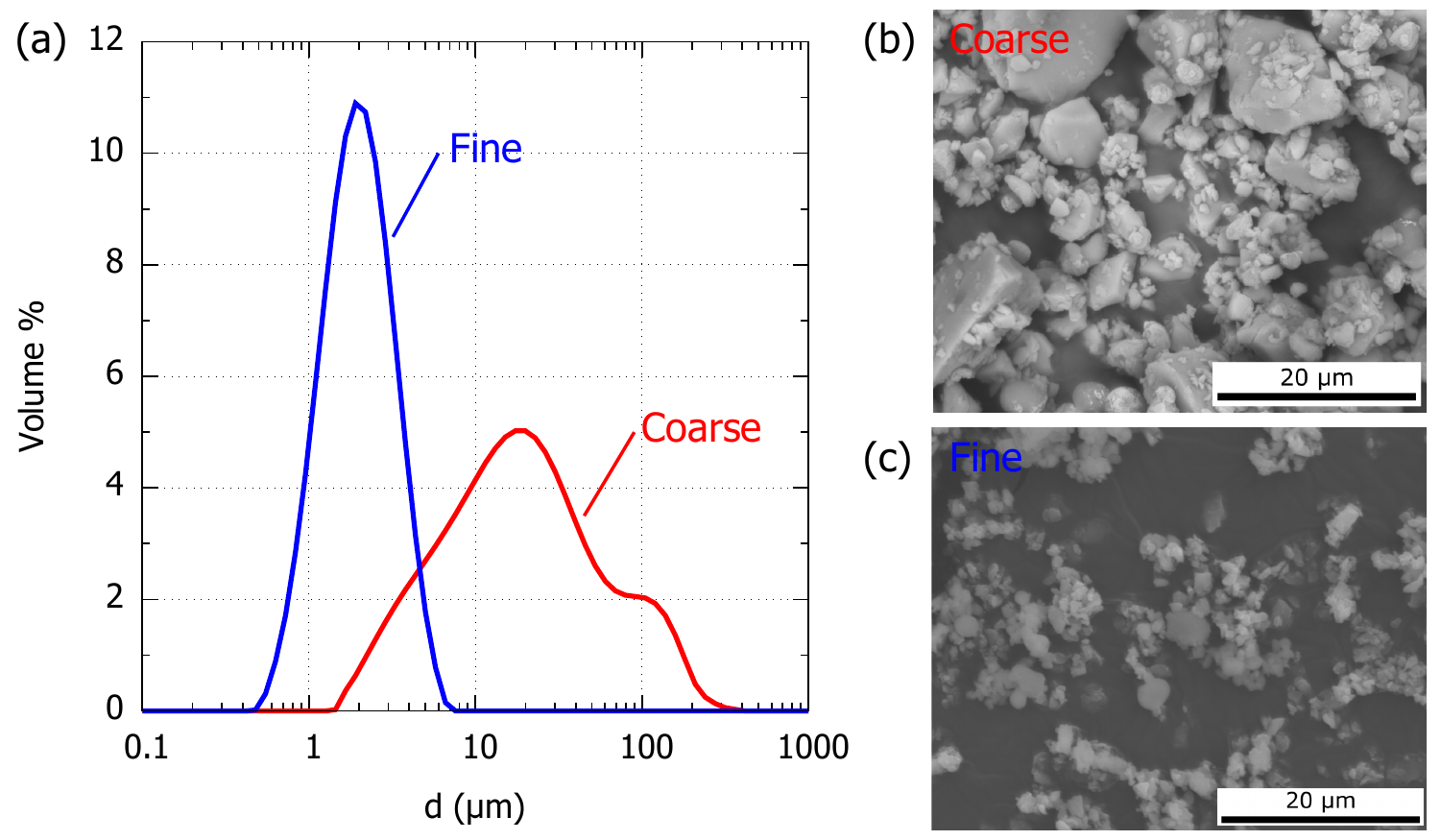}
		\caption{(a) Particle size distributions for coarse milled (planetary mill) and fine milled (attritor mill) phosphate glass, along with representative SEM images of the coarse milled (b) and fine milled (c) glass.}
		\label{fig:ParticleSize}
	\end{figure}

	The particle size distributions (determined on a volume basis by laser diffraction - Malvern Mastersizer 2000) of the coarse and fine milled glass powder are shown in Fig. \ref{fig:ParticleSize}. Coarse milling resulted in a median diameter ($d_{0.5}$) of 17 \textmu m (with $d_{0.1}$ = 4.0 \textmu m, $d_{0.9}$ = 85 \textmu m, $d_{max}$ = 360 \textmu m), after fine milling this was reduced to 1.6 \textmu m (with $d_{0.1}$ = 0.78 \textmu m, $d_{0.9}$ = 3.3 \textmu m, $d_{max}$ = 7.4 \textmu m). This is consistent with the particle size observed in the corresponding SEM images. The particle size reduction seen is in accord with what would be expected from a planetary mill for coarse milling, and the finer particles resulting from high-energy attritor milling \cite{Balaz2008, Suryanarayana2001}.
	
	\subsection{Composite characterisation}
	
\begin{figure}[h]
		\centering
		\includegraphics[width=9cm]{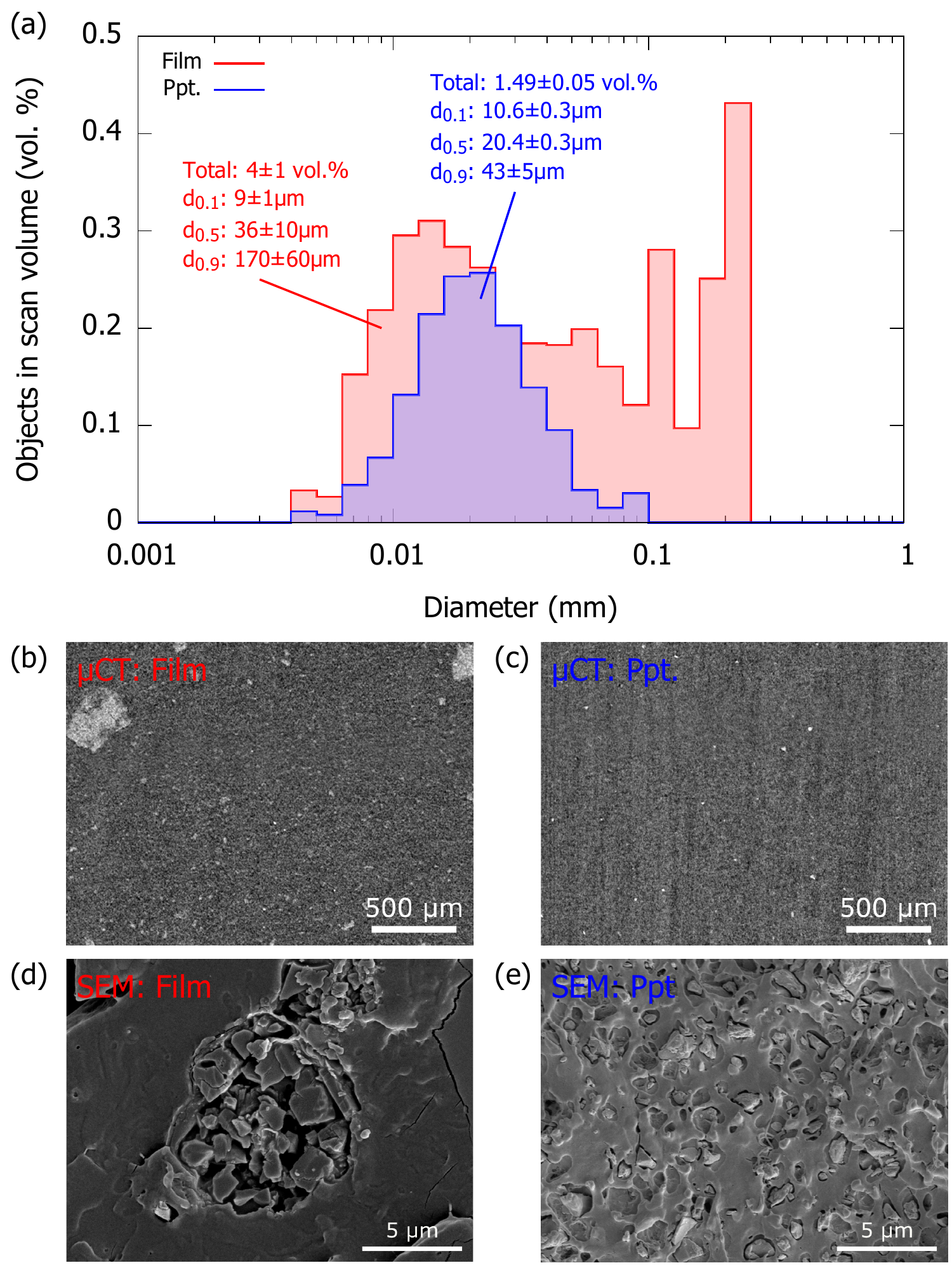}
		\caption{(a) Size distribution (by volume) of objects detected in \textmu CT scans, normalised to total scan volume, for PLLA with 30 wt.\% (17 vol.\%) phosphate glass, moulded from either composite films or composite precipitate (ppt.). (b, c) \textmu CT cross-sectional reconstructed images of these composite samples moulded from composite films (b) and composite precipitate (c). (d, e) SEM images of these composite samples moulded from composite films (d) and composite precipitate (e).}
		\label{fig:FillerDist}
	\end{figure}

	Two different methods were investigated for production of polymer-glass composites. The composite film casting process was used to produce thin composite films, followed by micro-injection moulding to fabricate composite samples for subsequent testing. The composite precipitation process was used to produce precipitated composite powder, again followed by micro-injection moulding to fabricate composite samples for later testing. Representative \textmu CT images are shown in Fig. \ref{fig:FillerDist}b-c of composite samples produced by moulding of composite films and composite precipitate. Both large and small agglomerates were visible in the samples moulded from composite films, while those moulded from composite precipitate showed a smaller amount and size of agglomerates.
	
	To quantify this difference 3D individual object analysis was used, to detect and measure individual agglomerate particles. These results are summarised in the distributions in Fig. \ref{fig:FillerDist}a. Objects detected by \textmu CT were converted to a volume distribution in order to allow comparison with the original particle size distributions in Fig. \ref{fig:ParticleSize}. Volume percentages were normalised to the total scan volume so that the amount of detected agglomerates could also be assessed. The samples tested were PLLA with a high glass content of 30 wt.\%, equivalent to 17 vol.\%, which have a tendency to agglomerate due to the high loading as seen in previous research \cite{Liu2002, Bleach2002}. In both cases the total vol.\% of agglomerates detected (1.49 - 4\%) was significantly lower than the actual glass content (17\%). This was due to the individual particle size, as detailed above, being comparable to the pixel size (1.49 \textmu m) - for accurate measurement a feature to pixel size ratio of 3:1 or greater is preferable \cite{DuPlessis2018, Bouxsein2010}. Therefore, well dispersed particles were not detected. The total vol.\% of objects detected could therefore be used to assess the amount of glass forming agglomerates - for samples moulded from composite films, this was 4\% of the total volume, so 24\% of the phosphate glass. By contrast for samples moulded from composite precipitate, only 1.49\% of volume was taken up by agglomerated glass, about 9\% of the total glass - a clear reduction in agglomeration. This is likely to be a slight overestimate however, since all particles detected by \textmu CT (minimum diameter of 4.5 \textmu m) were assumed to be glass agglomerates. Given the glass particle size, particularly the maximum particle size of 7.4 \textmu m, it is clear that a small portion of these will in fact be single particles rather than agglomerates.
	
	It is also important to assess not just the amount of agglomeration occurring, but also the size of these agglomerates. As the size distributions in Fig. \ref{fig:FillerDist}a show, samples moulded from composite precipitates showed a significant reduction in the amount of large agglomerates, with a reduction in the $d_{0.9}$ value from 170 $\pm 60$ \textmu m to 43 $\pm 5$ \textmu m. Composites moulded from films showed large amounts of agglomerates in the size range above 100 \textmu m, while those moulded from composite precipitate completely eliminated agglomerates above 100 \textmu m, with the bulk of agglomerates around 20 \textmu m. SEM results also show similar results to those observed by \textmu CT. Composites produced by moulding of cast films displayed large agglomerates of glass particles, as seen in Fig. \ref{fig:FillerDist}d. Conversely, composites fabricated by moulding of composite precipitate showed even dispersion of glass particles throughout the polymer matrix, with few large agglomerates seen.
	
	The composite precipitation method was able to achieve good dispersion of phosphate glass particles within the polymer matrix due to the fast formation of solid polymer on the surface of the glass particles. The glass particles were well dispersed in the slurry/dissolved polymer after sonication, and rather than transferring this to petri dishes for slow evaporation, ethanol was added to precipitate the polymer immediately. Once the solvent power of the solution was reduced enough for polymer precipitate to form, it nucleated on glass particles within the solution. Therefore, individual glass particles should be well covered by polymer precipitate, preventing agglomeration. This novel method allows production of well dispersed composites, avoiding the agglomeration that can be experienced when using solvent casting. It provides a useful alternative to melt blending, where the two components are mixed with the polymer in its molten state \cite{Wilberforce2011a}. This can result in significant polymer degradation and molecular weight reduction, owing to the extended processing time and high shear forces at high temperature \cite{Boccaccini2016, Odell1986}.
	
	\subsection{Mechanical testing}
	
\begin{figure}[h]
		\centering
		\includegraphics[width=12cm]{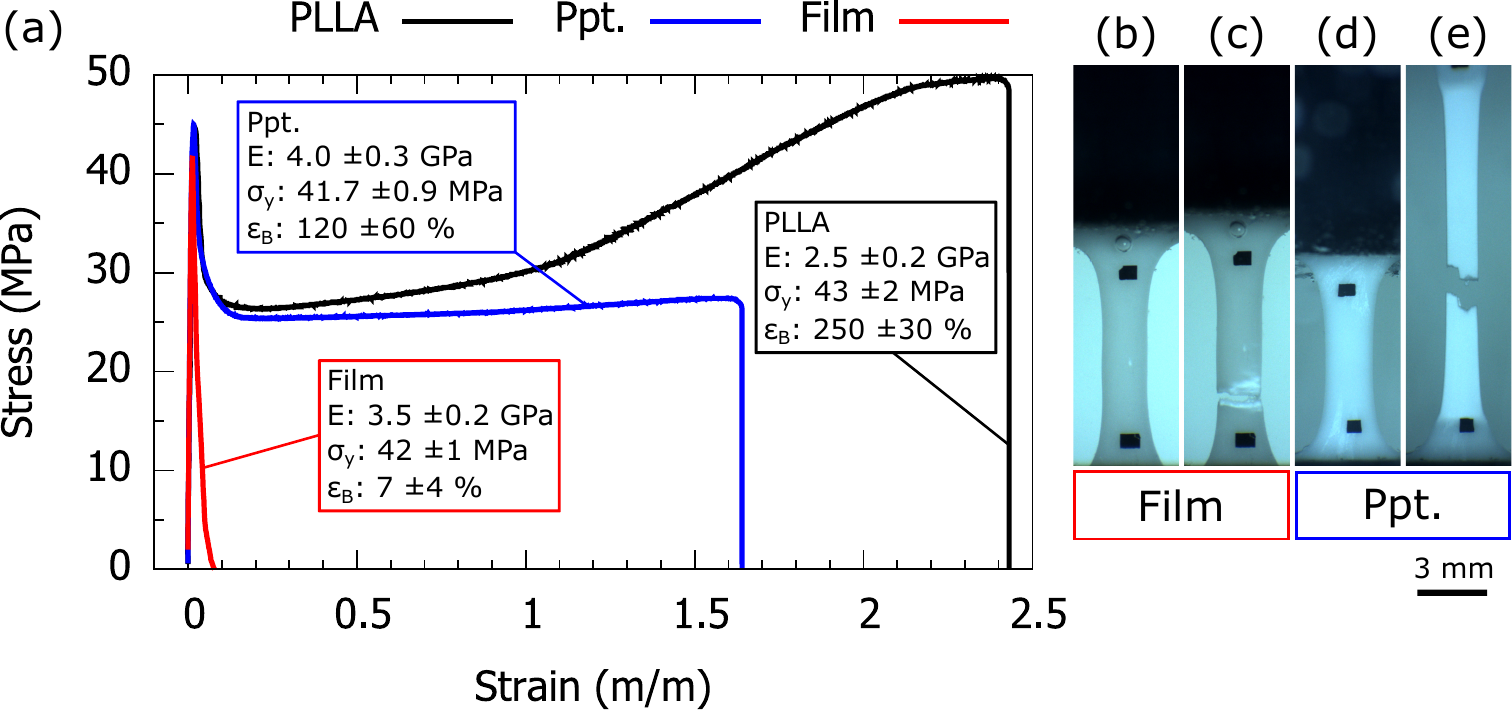}
		\caption{Typical mechanical testing results of composite (PLLA with 30wt.\% glass) samples moulded from composite films and composite precipitate, with neat PLLA for comparison, all tested in deionised water at 37\textcelsius, showing stress-strain curves (a). Photographs of samples are also shown: samples moulded from composite films, before testing (b) and after tensile failure (c), and samples moulded from composite precipitate, before testing (d) and after tensile failure (e).}
		\label{fig:Mech}
	\end{figure}

	In order to assess the impact of these two different production methods, and in particular the effect of agglomeration, mechanical testing was carried out using composites of PLLA with 30 wt.\% phosphate glass. These tests were carried out with the samples immersed in 37\textcelsius\ water, to simulate conditions within the body. For neat PLLA this resulted in lower yield strength and higher ductility than would be expected in ambient conditions, due to the combined effect of temperature and hydration as seen previously \cite{Oosterbeek2019}. Mechanical testing results are shown in Fig. \ref{fig:Mech}, where stress-strain curves can be seen, along with photographs of samples before and after tensile testing. Composite samples moulded from composite films and composite precipitate showed comparable strength and stiffness, with little significant differences seen between the measured $E$ and $\sigma_{y}$ for these two methods. The elastic modulus of the composite was unchanged as this is typically a result of the amount of glass present rather than the particle size \cite{Counto1964, Fu2008}, and the loading was kept constant at 30 wt.\%. Similarly, in the absence of strong interfacial bonding between the polymer and glass components of the composite, the agglomeration of glass particles would not be expected to significantly alter the yield strength, provided that agglomerates are not larger than the critical size for brittle failure by stress concentration \cite{Fu2008}.
	
	There was a large difference in the ductility of these materials, with composites moulded from films failing at 7\% ($\pm$4) strain, while those moulded from composite precipitate showed extensive plastic deformation, eventually failing at 120\% ($\pm$60) strain. These differences are also clear in photographs (Fig. \ref{fig:Mech}b-e), where the samples moulded from composite films showed tearing and crack propagation, while the samples moulded from composite precipitate showed necking and drawing before failure. The composites with more evenly dispersed glass (precipitation method) clearly displayed a much greater toughness. One crucial toughening mechanism for inorganic particle reinforced polymer composites is the debonding of the matrix from the particle. In composites where significant agglomeration has taken place, the surface area for energy absorption by debonding will be significantly lower, reducing the toughness. In addition, the greater stress concentrating effect of the large agglomerate, as well as the weak inter-particle bonding within the agglomerate, may all contribute to the lower ductility of the composites produced from moulding of films \cite{Fu2008, Liu2002}.
	
	\section{Conclusions}
	
	These results demonstrate the effectiveness of a method for production of polymer-glass composites. Phosphate glass particles with $d_{0.5} = 1.4$ \textmu m were produced and incorporated into well dispersed polymer-glass composites using a precipitation method followed by micro-injection moulding. Composites produced using this method demonstrated reduction in agglomerate size and amount, with $d_{0.9}$ reduced from 170 $\pm 60$ \textmu m to 43 $\pm 5$ \textmu m compared with composites produced using solvent film casting.
	
	This method requires straightforward modifications to common film casting procedures, and allows greater ductility to be achieved without sacrificing other mechanical properties. The method is versatile and adaptable, and applicable to most combinations of polymer and filler particles. It requires that suitable solvents be selected to ensure miscibility of the slurry and dissolved polymer without dissolution of the filler particles, and precipitation of the polymer once the final solvent is added.
	
	\section*{Acknowledgements}
	The authors thank Lucideon Ltd. (Stoke-on-Trent, United Kingdom) for providing materials and financial support of the project, as well as Mr Wayne Skelton-Hough and Mr Andrew Rayment for their technical support. RNO would also like to thank the Woolf Fisher Trust (Auckland, New Zealand) and the Cambridge Trust (Cambridge, United Kingdom), for provision of a PhD scholarship. Original data for this paper can be found at \url{https://doi.org/10.17863/CAM.63232}.
	
	\bibliographystyle{elsarticle-num}
	
	\bibliography{Refs}

	\vspace{3cm}
	
	\begin{flushleft}
		
		Published journal article:\\
		\url{https://doi.org/10.1016/j.jmbbm.2021.104767}\\
		\vspace{0.5cm}
		Copyright \textcopyright\ \href{https://creativecommons.org/licenses/by-nc-nd/2.0/uk/}{CC-BY-NC-ND}\\
		\vspace{0.2cm}
		\includegraphics[width=0.2\linewidth]{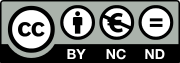}	
		
	\end{flushleft}

\end{document}